\documentclass[a4paper,10pt,twoside]{cpc-hepnp}

\usepackage{multicol}
\usepackage{graphicx}
\usepackage{booktabs}
\usepackage{amssymb,bm,mathrsfs,bbm,amscd}
\usepackage[tbtags]{amsmath}
\usepackage{lastpage}
\usepackage{CJK}

\begin{document}
\begin{CJK*}{GB}{gbsn}

\fancyhead[c]{\small Chinese Physics C~~~Vol. xx, No. x (2017)
xxxxxx} \fancyfoot[C]{\small xxxxxx-\thepage}

\footnotetext[0]{Received 25 August 2017}

\title{Microscopic analysis of octupole shape transitions in neutron-rich actinides with relativistic energy density functional\thanks{Supported by National Natural Science Foundation of China (11475140, 11575148) }}

\author{%
      Zhong Xu%
\quad Zhi-Pan Li\email{zpliphy@swu.edu.cn}%
}
\maketitle

\address{%
School of Physical Science and Technology, Southwest University, Chongqing 400715, China\\
}

\begin{abstract}
Quadrupole and octupole deformation energy surfaces, low-energy excitation spectra, and electric transition rates in eight neutron-rich isotopic chains --  Ra, Th, U, Pu, Cm, Cf, Fm, and No -- are systematically analyzed using a quadrupole-octupole collective Hamiltonian model, with parameters determined by constrained reflection-asymmetric and axially-symmetric relativistic mean-field calculations based on the PC-PK1 energy density functional. The theoretical results of low-lying negative-parity bands, odd-even staggering, average octupole deformations $\langle\beta_3\rangle$, and  $B(E3; 3^-_1\to 0^+_1)$ show evidence of a shape transition from nearly spherical to stable octupole-deformed, and finally octupole-soft equilibrium shapes in the neutron-rich actinides.  A microscopic mechanism for the onset of stable octupole deformation is also discussed in terms of the evolution of single-nucleon orbitals with deformation.
\end{abstract}



\footnotetext[0]{\hspace*{-3mm}\raisebox{0.3ex}{$\scriptstyle\copyright$}2013
Chinese Physical Society and the Institute of High Energy Physics
of the Chinese Academy of Sciences and the Institute
of Modern Physics of the Chinese Academy of Sciences and IOP Publishing Ltd}%

\begin{multicols}{2}

\section{Introduction}

The study of octupole-deformed (reflection asymmetric) shapes and shape transitions presents a recurrent theme in nuclear structure physics.  Octupole-deformed  shapes are characterized by the presence of low-lying negative-parity bands, and by pronounced electric octupole transitions \cite{Butler96,Ahmad93,BW.15,Butler16}.  In the case of static octupole deformation, for instance, the lowest positive-parity even-spin states and the negative-parity odd-spin states form an alternating-parity band, with states connected by the enhanced $E1$ transitions. Recently, evidence for pronounced octupole deformation in  $^{224}$Ra \cite{Gaff13}, $^{144}$Ba \cite{Bucher16}, and $^{146}$Ba \cite{Bucher17} has been reported in Coulomb excitation experiments with radioactive ion beams. The renewed interest in studies of reflection asymmetric nuclear shapes using accelerated radioactive beams point to the importance of a timely systematic theoretical analysis of quadrupole-octupole collective states of nuclei in different mass regions.

A series of theoretical models have been applied to the studies of octupole-deformed shapes and the evolution of the corresponding negative-parity collective states, including the energy density functionals or their simplest realization: self-consistent mean-field models \cite{Bonche86,Bonche88,Egido91,Rutz95,Geng07,Guo10,Robledo10,Robledo11,Rodr12,Robledo13,Robledo15,Zhao12,ZhouSG16,Zhao17,Nomura13,Nomura14,Nomura15,Agbe16,Agbe17,Zhang10,Zhang10b,Li13,Li16,Yao15,Zhou16,Xia17,Ebata17,Bernard16}, macroscopic+microscopic (MM) models \cite{Naza84,Moller08,Wang15}, algebraic (or interacting boson) models \cite{Scho78,Otsuka88}, phenomenological collective models \cite{Bizzeti04,Bonatsos05,Bizzeti13,Minkov12,Jolos12}, and the reflection asymmetric shell model \cite{Chen15}.

In particular, nuclear energy density functionals (EDFs) enable a complete and accurate description of ground-state properties and collective excitations over the entire chart of nuclides \cite{Bender03,VALR05,Meng06,Stone07,Nik11,Meng16}. Both non-relativistic and relativistic EDFs have successfully been applied to the description of the evolution of single-nucleon shell structures and related nuclear shapes and shape transitions. To compute excitation spectra and transition rates, however, the EDF framework has to be extended to take into account the restoration of symmetries broken in the mean-field approximation, and fluctuations in the collective coordinates. A straightforward approach is the generator coordinate method (GCM) combined with projection techniques, and recently it has been implemented for octupole-deformed shapes, based on both nonrelativistic \cite{Bernard16} and relativistic \cite{Yao15,Zhou16} EDFs. Using this method, however, it is rather difficult to perform a systematic study of low-lying quadrupole and octupole states in different mass regions, because GCM is very time-consuming for heavy systems.  An alternative approach is the EDF-based quadrupole-octupole collective Hamiltonian (QOCH) \cite{Li13,Li16,Xia17}. The collective Hamiltonian can be derived from the GCM in the Gaussian overlap approximation \cite{Ring80}, and the validity of this approximate method was recently demonstrated in a comparison with a full GCM calculation for the shape coexisting nucleus $^{76}$Kr \cite{Yao14}.

Recently, we have applied the EDF-based QOCH to a systematic analysis of spectroscopy of quadrupole and octupole states in fourteen isotopic chains: Xe, Ba, Ce, Nd, Sm, Gd, Rn, Ra, Th, U, Pu, Cm, Cf, and Fm. The microscopic QOCH model, based on the PC-PK1 energy density functional \cite{Zhao10}, is shown to accurately describe the empirical trend of low-energy quadrupole and octupole collective states, and the predicted spectroscopic properties are also consistent with recent microscopic calculations based on both relativistic and non-relativistic energy density functionals. The resulting low-energy negative-parity bands, average octupole deformations, and transition rates show evidence for octupole collectivity in  mass regions centered at both $Z\sim58, N\sim90$ and $Z\sim90, N\sim136$. The success of the EDF-based QOCH model in these mass regions enables us to search for the next possible octupole-deformed mass region by analyzing both deformation energy surfaces and low-lying spectroscopy.

Very recently, a systematic search for axial octupole deformation in the actinides and superheavy nuclei with proton numbers $Z=88-126$ and neutron numbers from the two-proton drip line up to $N=210$ was performed using the mean-field framework of relativistic density functional theory, and octupole-deformed minima were predicted in the nuclei around $Z\sim96, N\sim196$. Therefore, in this study we employ the  EDF-based QOCH to perform a systematic calculation of even-even neutron-rich heavy nuclei ($88\leq Z\leq 102$ and $190\leq N\leq212$). Low-energy spectra and transition rates for both positive- and negative-parity states of 96 nuclei are calculated using the QOCH with parameters determined by self-consistent reflection-asymmetric relativistic mean-field calculations based on the PC-PK1 energy density functional \cite{Zhao10}. The relativistic functional PC-PK1 was adjusted to the experimental masses of a set of 60 spherical nuclei along isotopic or isotonic chains, and to the charge radii of 17 spherical nuclei.  PC-PK1 has been successfully employed in studies of nuclear masses \cite{Zhang14,Lu15}, and spectroscopy of low-lying quadrupole states \cite{Quan17}.

The article is organized as follows. Section {\ref{ScII}} presents a brief review of the EDF-based QOCH model. The systematics of collective deformation energy surfaces, excitation energies of low-lying positive- and negative-parity states, odd-even staggering, electric dipole, quadrupole, and octupole transition rates, calculated with the QOCH model, are discussed in Section {\ref{ScIII}}. Section {\ref{IV}} contains a summary and concluding remarks.

%
\section{\label{ScII}Theoretical Framework}
%
 %

Detailed formalism of the quadrupole-octupole collective Hamiltonian has been presented in Refs. \cite{Li16,Xia17}. In this section, for completeness, a brief introduction is presented. The QOCH, which can simultaneously treat the axially quadrupole-octupole vibrational and rotational excitations, is expressed in terms of two deformation parameters $\beta_2$ and $\beta_3$, and three Euler angles $\left( \phi ,\theta ,\psi  \right)\equiv \Omega $ that define the orientation of the intrinsic principal axes in the laboratory frame,
\begin{equation}
\begin{split}
{{\hat{H}}_{\text{coll}}}=&-\dfrac{\hbar^2}{2\sqrt{w\mathcal{I}}}
          \left[\dfrac{\partial}{\partial \beta_2}\sqrt{\dfrac{\mathcal{I}}{w}}B_{33}\dfrac{\partial}{\partial \beta_2}
         -\dfrac{\partial}{\partial \beta_2}\sqrt{\dfrac{\mathcal{I}}{w}}B_{23}\dfrac{\partial}{\partial \beta_3}\right. \\
       &\left.-\dfrac{\partial}{\partial \beta_3}\sqrt{\dfrac{\mathcal{I}}{w}}B_{23}\dfrac{\partial}{\partial \beta_2}
         +\dfrac{\partial}{\partial \beta_3}\sqrt{\dfrac{\mathcal{I}}{w}}B_{22}\dfrac{\partial}{\partial \beta_3}\right] \\
       &+\dfrac{\hat{J}^2}{2\mathcal{I}}+{{V}_{\text{coll}}}( {{\beta }_{2}}, {{\beta }_{3}} ).
\end{split}
   \label{eq:CH}
\end{equation}
$\hat{J}$ denotes the component of angular momentum perpendicular to the symmetric axis in the body-fixed frame of a nucleus. The mass parameters $B_{22}$, $B_{23}$, and $B_{33}$, the moments of inertia $\mathcal{I}$, and collective potential $V_\text{coll}$ depend on the quadrupole and octupole deformation variables $\beta_2$ and $\beta_3$. The additional quantities that appear in the vibrational kinetic energy, $w=\sqrt{B_{22}B_{33}-B_{23}^2}$, determine the volume element in the collective space
\begin{equation}
\int d\tau_\text{coll}=\int\sqrt{w\cal{I}}d\beta_2 d\beta_3 d\Omega.
\end{equation}

The eigenvalue problem of the collective Hamiltonian \eqref{eq:CH} is solved using an expansion of eigenfunctions in terms of a complete set of basis functions that depend on the collective coordinates $\beta_2$, $\beta_3$, and $\Omega$. The collective wave functions are thus obtained as
\begin{equation}
\Psi _{\alpha }^{IM\pi }\left( {{\beta }_{2}},{{\beta }_{3}},\Omega  \right)=\psi _{\alpha }^{I\pi }\left( {{\beta }_{2}},{{\beta }_{3}} \right)\left| IM0 \right\rangle.
  \label{eq:wave}
\end{equation}
The reduced $E\lambda$ values are calculated from the relation
\begin{equation}
B\left( E\lambda ,{{I}_{i}}\to {{I}_{f}} \right)={{\left\langle  {{I}_{i}}0\lambda 0 | {{I}_{f}}0 \right\rangle }^{2}}{{\left| \int{d{{\beta }_{2}}d{{\beta }_{3}}
\sqrt{w\mathcal{I}}{{\psi }_{i}}{\mathcal{M}_{E\lambda }}
\left( {{\beta }_{2}},{{\beta }_{3}} \right)\psi _{f}^{*}} \right|}^{2}},
  \label{eq:BE}
\end{equation}
where $\mathcal{M}_{E\lambda}(\beta_2, \beta_3)$ denotes the electric moment of order $\lambda$. In microscopic models it is calculated as $\langle\Phi(\beta_2, \beta_3)|\hat{\mathcal{M}}(E\lambda)|\Phi(\beta_2, \beta_3)\rangle$, where $\Phi(\beta_2, \beta_3)$ is the nuclear wave function.

In the framework of the EDF-based QOCH model, the collective parameters of QOCH in Eq. (\ref{eq:CH}) are all determined from the EDF microscopically. The moments of inertia are calculated according to the Inglis-Belyaev formula \cite{Inglis56,Belyaev61}:
\begin{equation}
\label{eq:MOI}
\mathcal{I} = \sum_{i,j}{\frac{\left(u_iv_j-v_iu_j \right)^2}{E_i+E_j}
  | \langle i |\hat{J} | j  \rangle |^2},
\end{equation}
where $\hat J$ is the angular momentum along the axis perpendicular to the symmetric axis,  and the summation runs over the proton and neutron quasiparticle states. The quasiparticle energies $E_i$, occupation probabilities $v_i$, and single-nucleon wave functions $\psi_i$ are determined by solutions of the constrained EDF. The mass parameters are calculated in the perturbative  cranking approximation \cite{Girod79,Xia17}
\begin{equation}
\label{eq:}
B_{\lambda\lambda^\prime}(q_2,q_3)=\frac{\hbar^2}{2}
 \left[\mathcal{M}_{(1)}^{-1} \mathcal{M}_{(3)} \mathcal{M}_{(1)}^{-1}\right]_{\lambda\lambda^\prime}\;,
\end{equation}
with
\begin{equation}
\label{masspar-M}
\mathcal{M}_{(n),\lambda\lambda^\prime}(q_2,q_3)=\sum_{i,j}
 {\frac{\left\langle i\right|\hat{Q}_{\lambda}\left| j\right\rangle
 \left\langle j\right|\hat{Q}_{\lambda^\prime}\left| i\right\rangle}
 {(E_i+E_j)^n}\left(u_i v_j+ v_i u_j \right)^2}\;,
\end{equation}
where $\hat Q_2$ and $\hat Q_3$ are the mass quadrupole and octupole operators, respectively, and $q_\lambda=\langle\hat Q_\lambda\rangle$.

The collective potential $V_{\rm coll}$ in Eq. (\ref{eq:CH}) is obtained by subtracting the vibrational and rotational zero-point energy (ZPE) corrections from the total mean-field energy:
\begin{equation}
\label{eq:Vcoll}
V_{\rm coll}(\beta_2, \beta_3)=E_{\rm MF}(\beta_2, \beta_3)-\Delta V_{\rm vib}(\beta_2, \beta_3)-\Delta V_{\rm rot}(\beta_2, \beta_3).
\end{equation}
The vibrational and rotational ZPE corrections are calculated in the cranking approximation \cite{Girod79}:
\begin{equation}
\label{ZPE-vib}
\Delta V_{\rm vib}(\beta_2, \beta_3) = \frac{1}{4}
\textnormal{Tr}\left[\mathcal{M}_{(3)}^{-1}\mathcal{M}_{(2)}  \right],
\end{equation}
and
\begin{equation}
\label{ZPE-rot}
\Delta V_{\rm rot}(\beta_2, \beta_3)=\frac{\langle\hat J^2\rangle}{2{\cal I}},
\end{equation}
respectively.

\section{\label{ScIII}Results and discussion}

The principal objective of this study is a systematic analysis that includes collective deformation energy surfaces (DESs), excitation energies and average quadrupole and octupole deformations of low-lying states, electric dipole, quadrupole, and octupole transitions for even-even neutron-rich heavy nuclei ($88\leq Z\leq 102$ and $190\leq N\leq212$) using the EDF-based QOCH model. To determine the collective input for the QOCH, we perform a constrained reflection-asymmetric relativistic mean-field plus BCS (RMF+BCS) calculation, with the effective interaction in the particle-hole channel defined by the relativistic density functional PC-PK1 \cite{Zhao10}, and a density independent $\delta$-force \cite{Bender00} in the particle-particle channel. The strength parameter of the $\delta$-force, 333.9 MeV fm$^3$ (397.0 MeV fm$^3$) for neutrons (protons), is determined to reproduce the corresponding pairing gap of the spherical configuration of $^{300}\text{Fm}$, calculated using the relativistic Hartree-Bogoliubov (RHB) model with the finite-size separable pairing force \cite{Nik14}. This can be done because the essential effects of the off-diagonal parts of the pairing field neglected in the RMF+BCS calculations can be recovered by simply renormalizing the pairing strength, and consequently the low-energy structure is in good agreement with the predictions of the RHB model \cite{Xiang13}.  Moreover, the RHB model with  finite-size separable pairing force was successfully used in the description of octupole deformations \cite{Agbe16} and low-energy excitation spectra \cite{Nomura14,Li16}. The solution of the single-nucleon Dirac equation in RMF+BCS is obtained by expanding the nucleon wave functions in an axially deformed harmonic oscillator basis with 20 major shells.

Figures \ref{RaTh-pes}, \ref{UPu-pes}, \ref{CmCf-pes}, and \ref{FmNo-pes} display the DESs of the even-even Ra, Th, U, Pu, Cm, Cf, Fm, and No isotopes in the $\beta_2$-$\beta_3$ plane, calculated with the RMF+BCS using the functional PC-PK1 and $\delta$-force pairing. The quadrupole and octupole deformations that correspond to the global minima, as well as the octupole deformation energies $\Delta E_\text{oct}$, defined as the energy differences between the non-octupole deformed minima ($\beta_3=0$) and the global minima, are also plotted in Fig. \ref{bmin}. The  equilibrium quadrupole deformations for all the isotopic chains increase gradually, from nearly spherical to well-deformed shapes, as the neutron number increases from 190 to 212.  All the isotopic chains except Ra exhibit a very interesting shape evolution: from nearly spherical to octupole-deformed, and finally octupole-soft equilibrium shapes. Stable equilibrium octupole deformations are calculated in $^{286-292}$Th, $^{286-294}$U, $^{286-294}$Pu, $^{288-296}$Cm, $^{290-298}$Cf, $^{292-302}$Fm, and $^{294-302}$No. There are peaks at $N\sim196$ for the octupole deformation energies $\Delta E_\text{oct}$ and the maximum value is $\sim1.8$ MeV, observed in $^{290}$Pu and $^{292}$Cm. For the Ra isotopic chain, weak octupole deformation is predicted in $^{284-294}$Ra but the energy surfaces are very shallow with respect to the octupole degree of freedom. Similar shape transitions in the actinides have also been obtained in studies based on different relativistic energy density functionals \cite{Agbe17}. Some differences between these calculations are found in the exact location of non-zero equilibrium octupole deformation and the corresponding octupole deformation energies. This can be attributed to the details of the single-particle spectra, especially levels with $\Delta j=3$ and $\Delta l=3$, and also to different treatment of pairing correlations \cite{Agbe17}.


\end{multicols}
\ruleup


\begin{center}
\includegraphics[width=12.8cm]{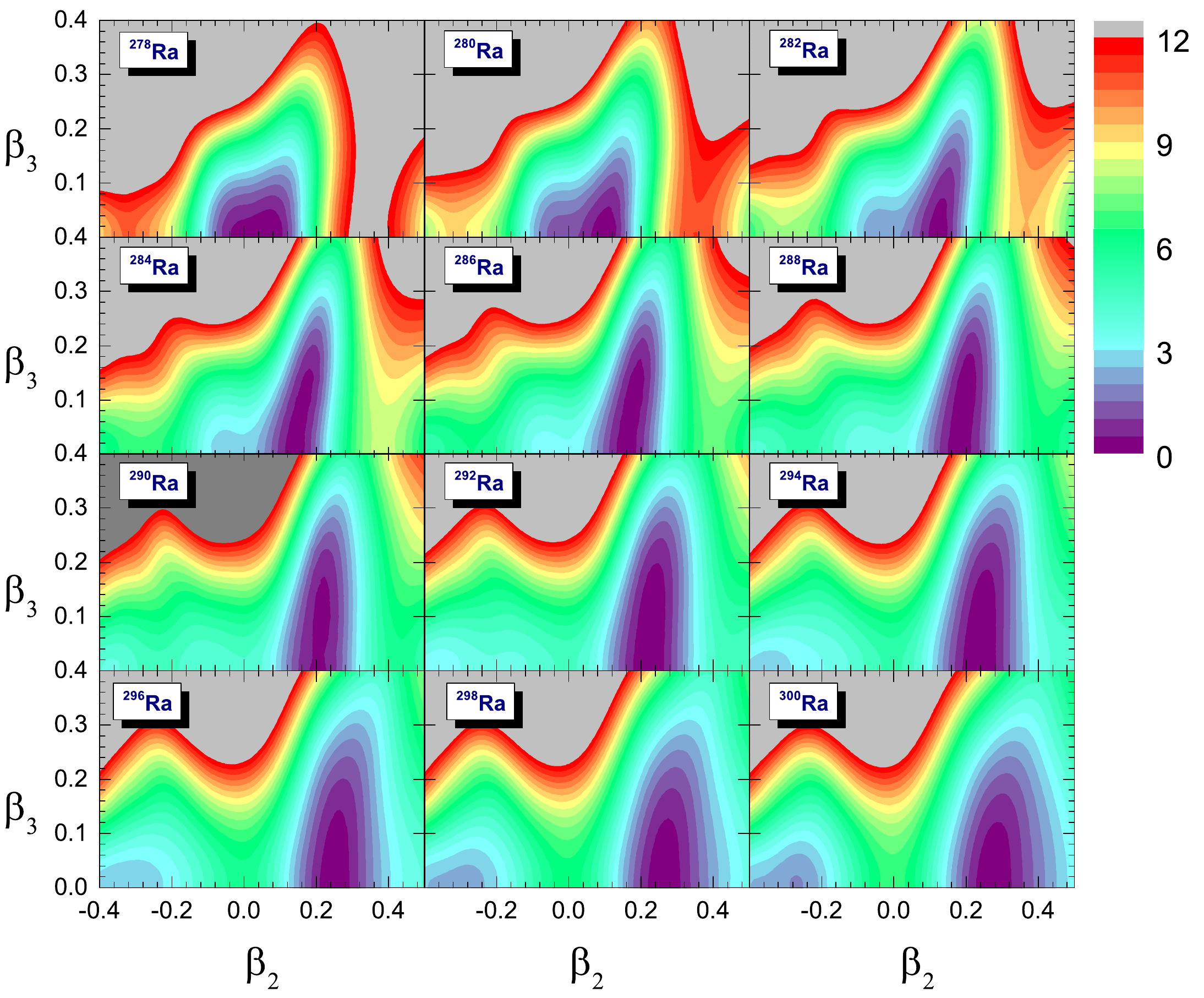}
\includegraphics[width=12.8cm]{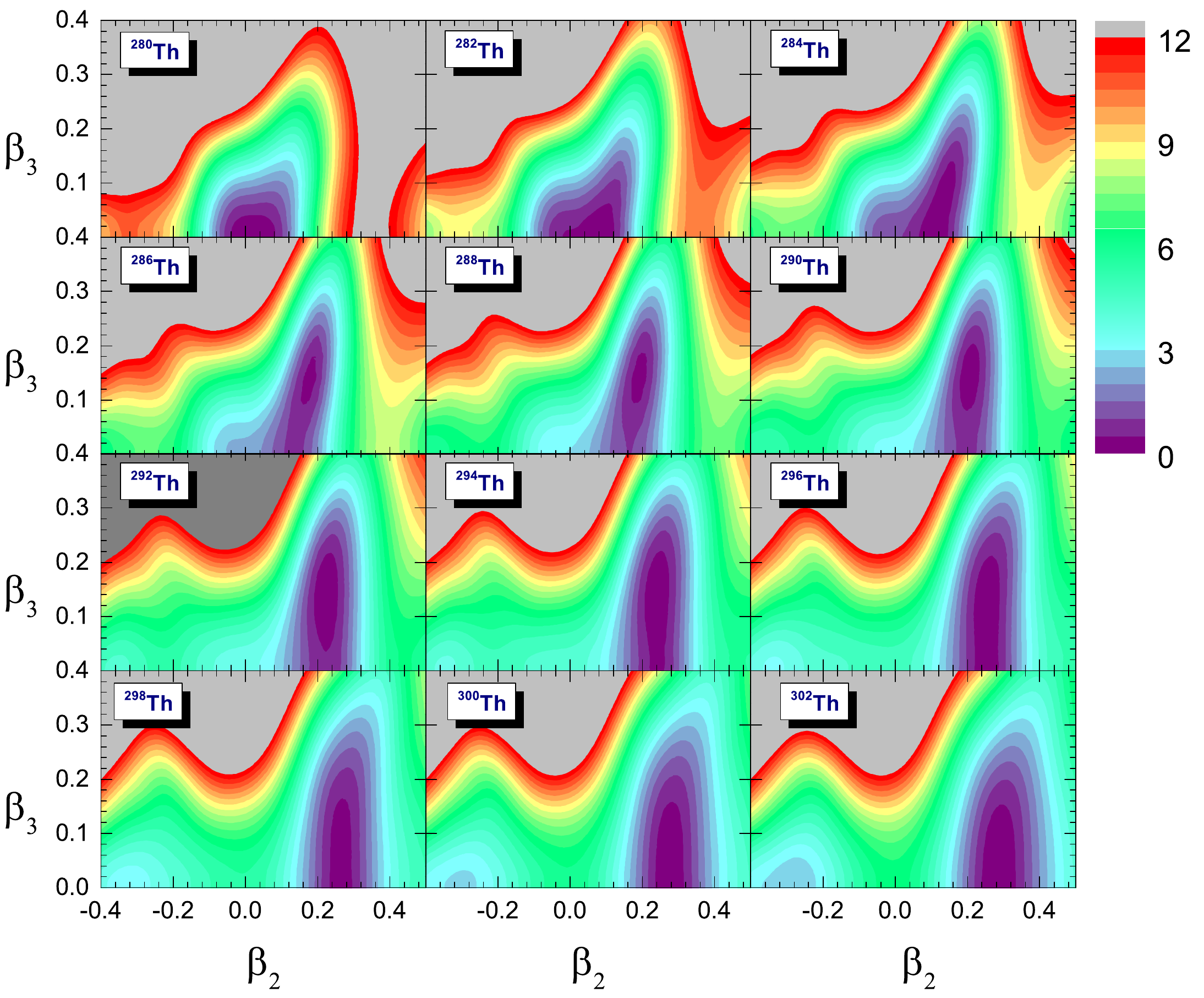}
\figcaption{\label{RaTh-pes} Axially-symmetric quadrupole-octupole deformation energy surfaces of $\text{Ra}$ and $\text{Th}$ isotopes in the $\beta_2$-$\beta_3$ plane calculated by self-consistent RMF+BCS. For each nucleus, energy values are normalized with respect to the energy minimum of the ground states. The contours join points on the surface with the same energy, and the separation between neighboring contours is 0.5 MeV.}
\end{center}

\begin{center}
\includegraphics[width=13cm]{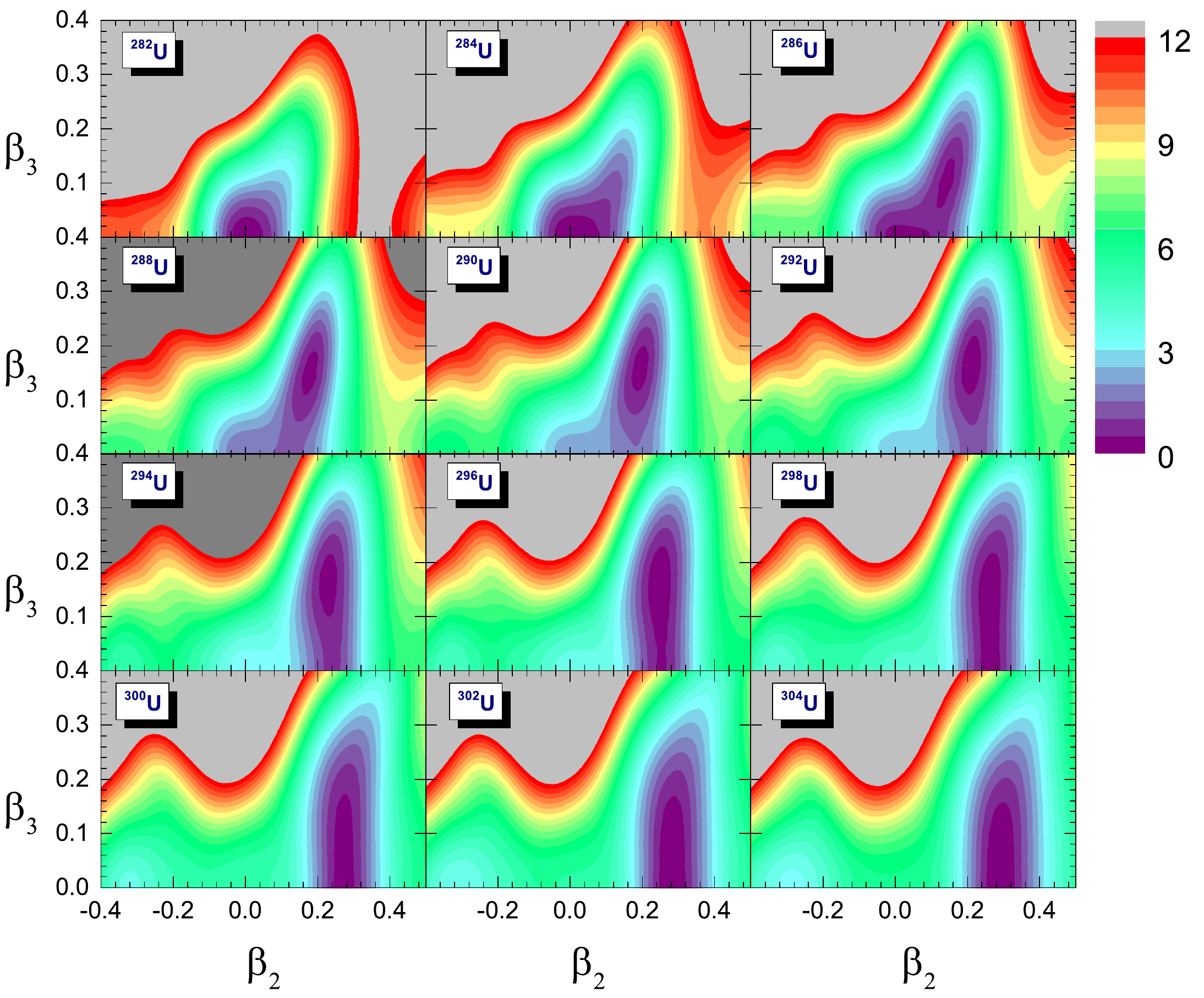}
\includegraphics[width=13cm]{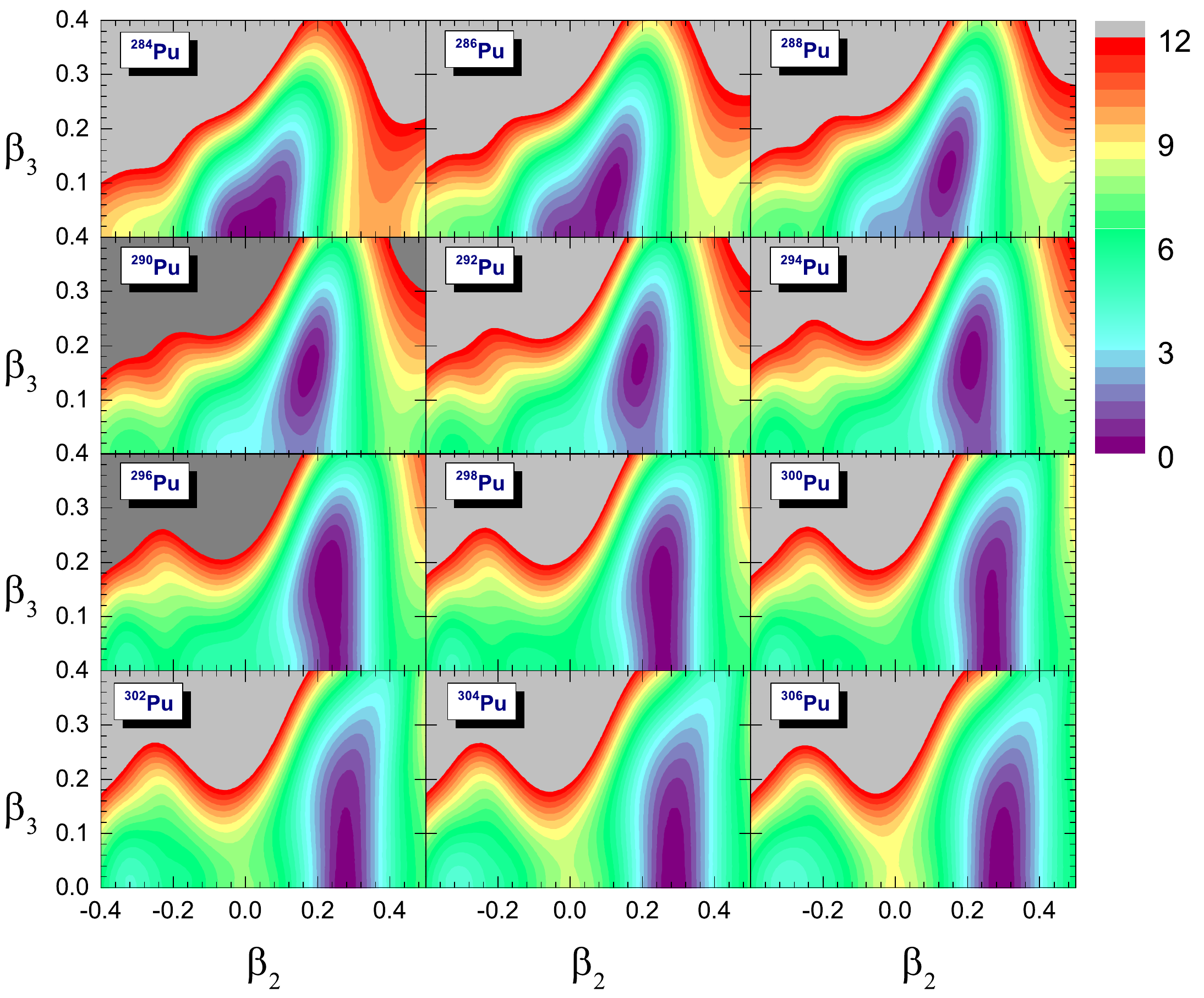}
\figcaption{\label{UPu-pes} Same as Fig. \ref{RaTh-pes} but for $\text{U}$ and $\text{Pu}$ isotopes.}
\end{center}

\begin{center}
\includegraphics[width=13cm]{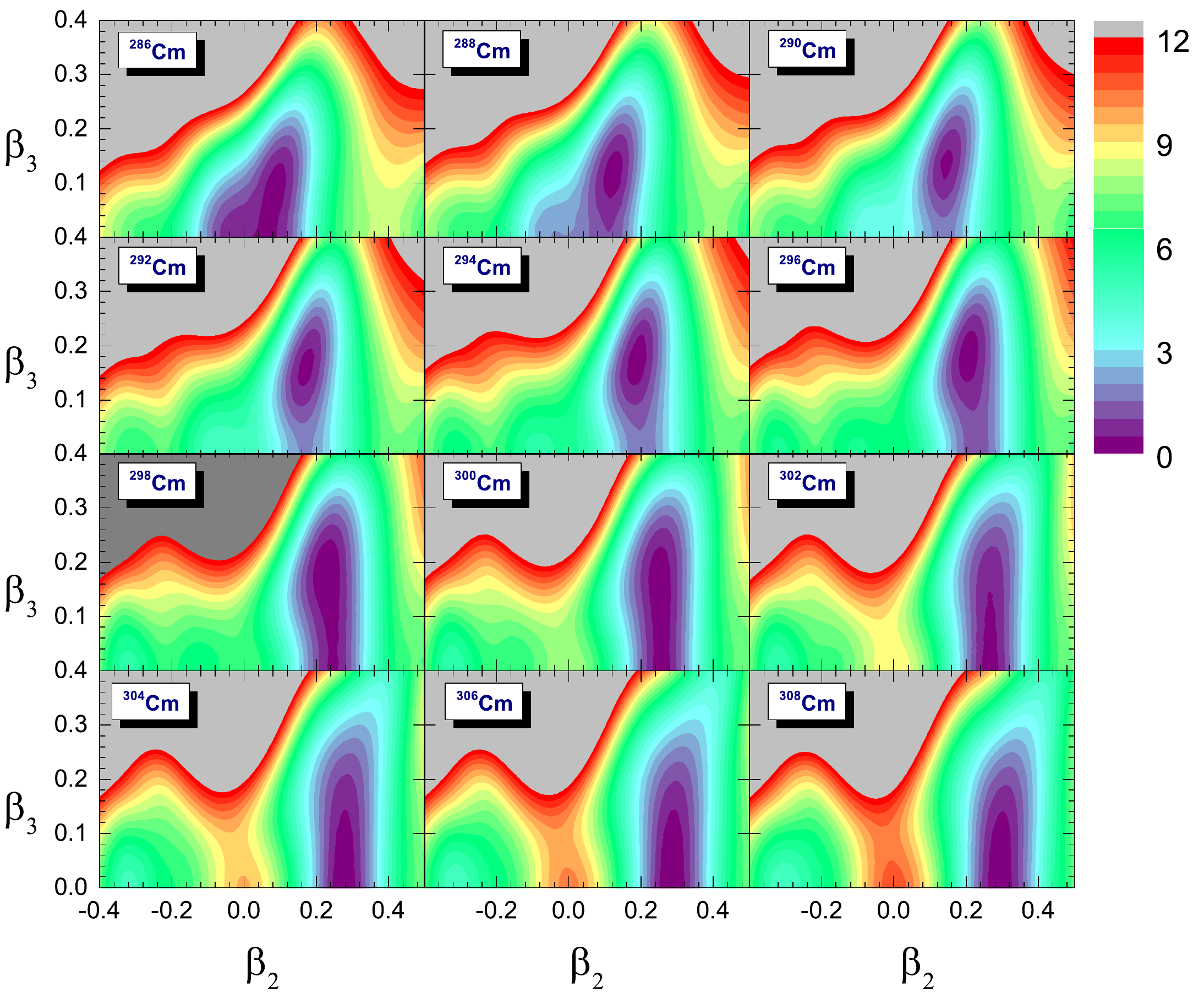}
\includegraphics[width=13cm]{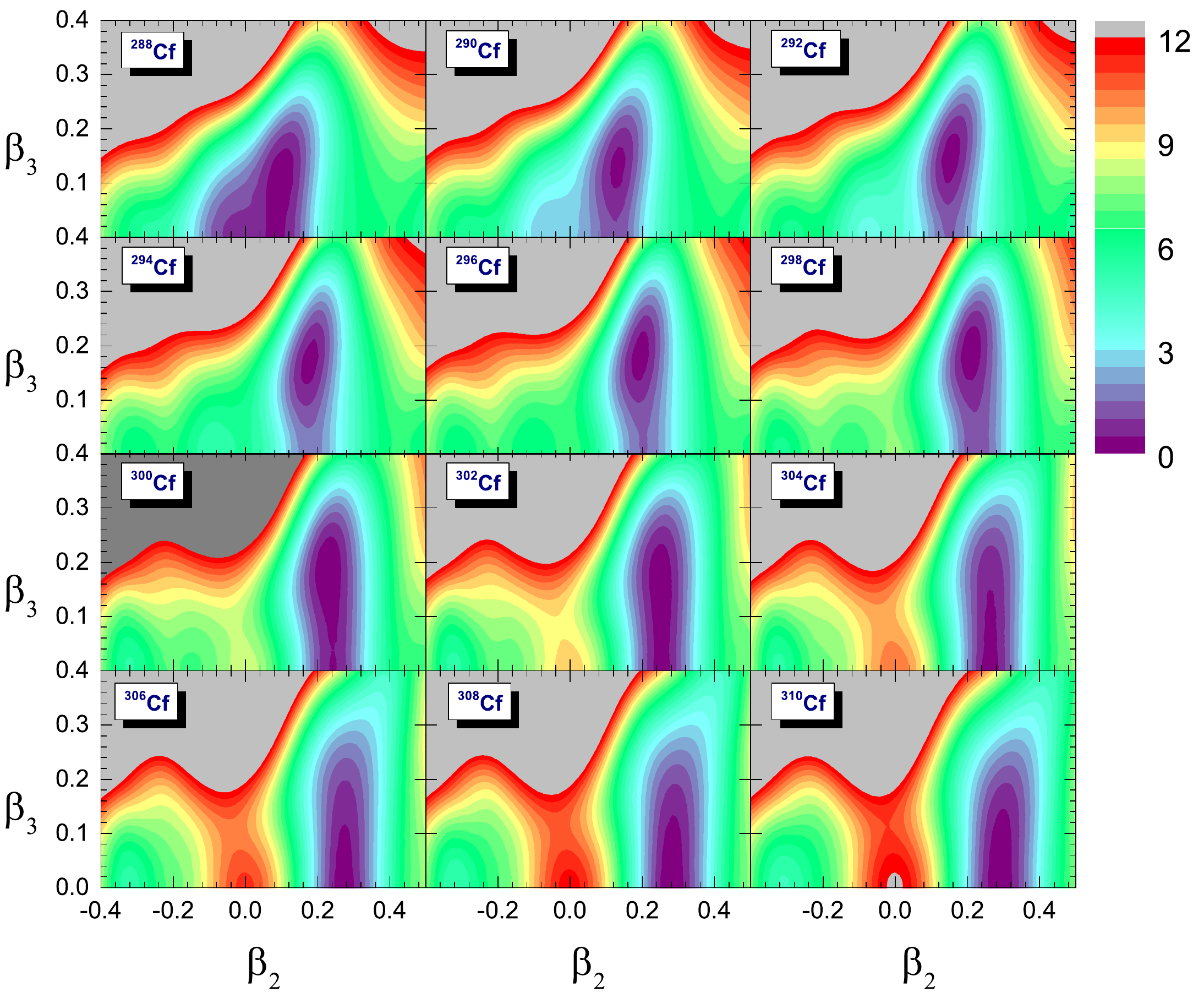}
\figcaption{\label{CmCf-pes} Same as Fig. \ref{RaTh-pes} but for $\text{Cm}$ and $\text{Cf}$ isotopes.}
\end{center}

\begin{center}
\includegraphics[width=13cm]{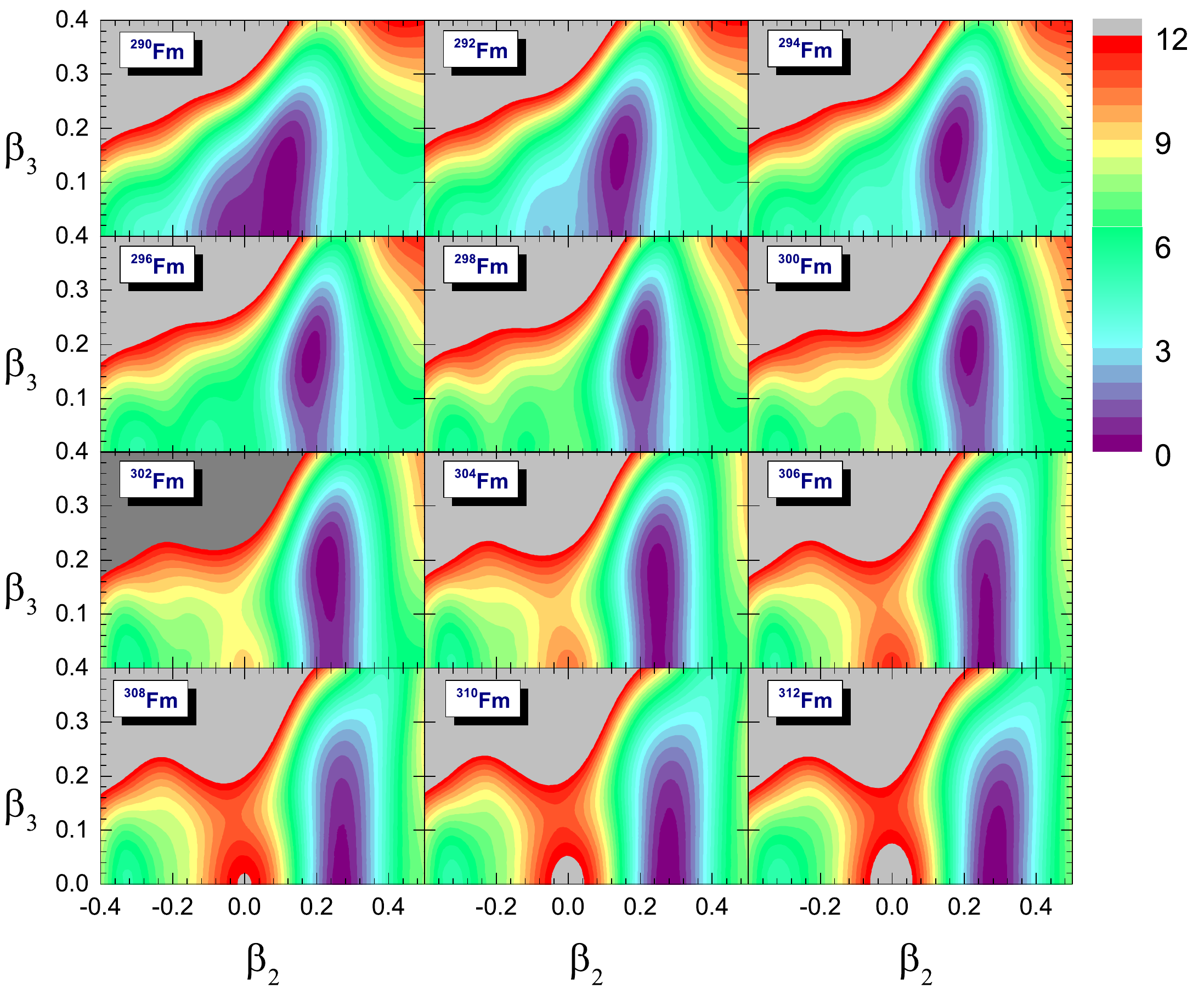}
\includegraphics[width=13cm]{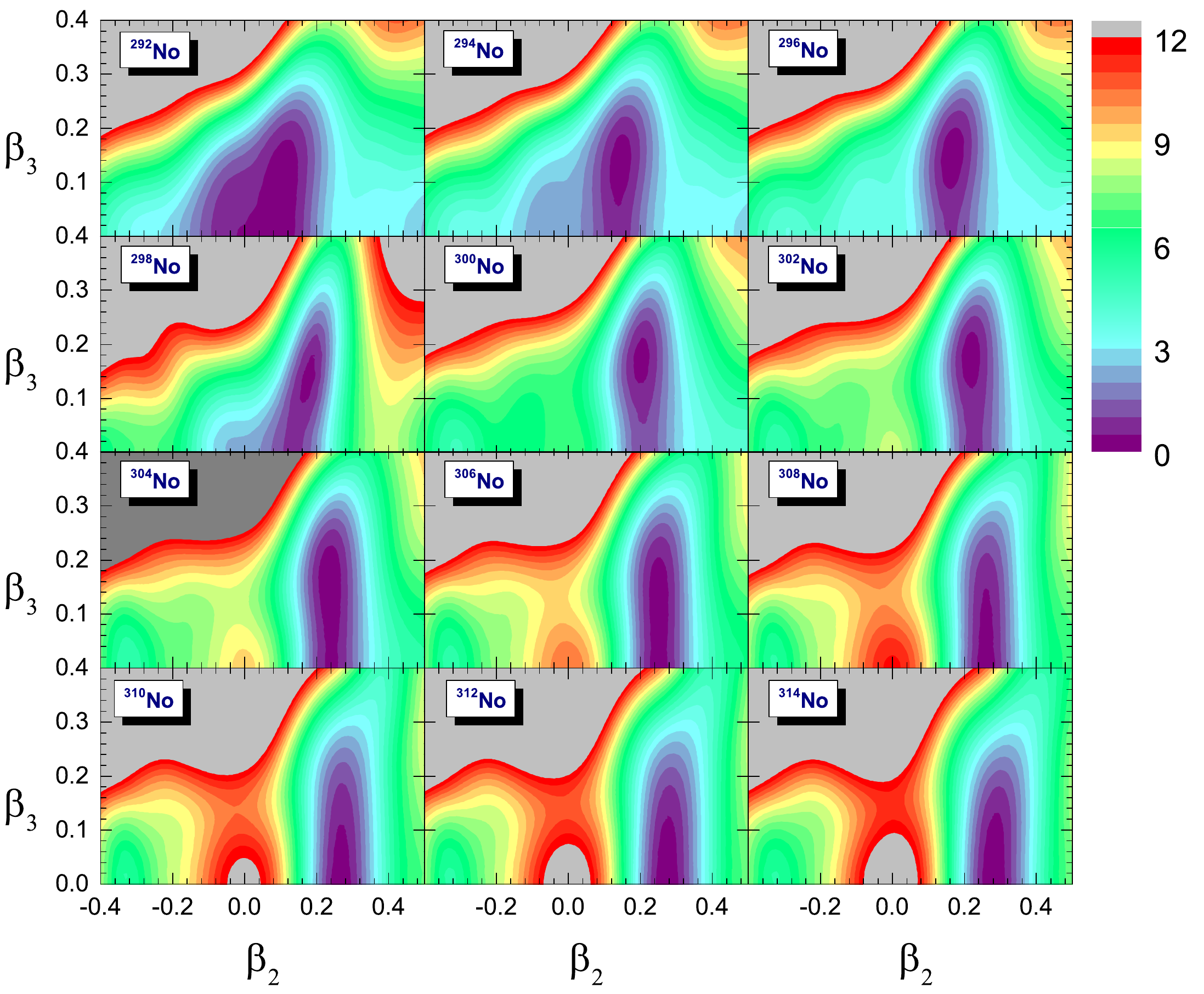}
\figcaption{\label{FmNo-pes} Same as Fig. \ref{RaTh-pes} but for $\text{Fm}$ and $\text{No}$ isotopes.}
\end{center}
\ruledown

\begin{multicols}{2}

\begin{center}
\includegraphics[width=7.5cm]{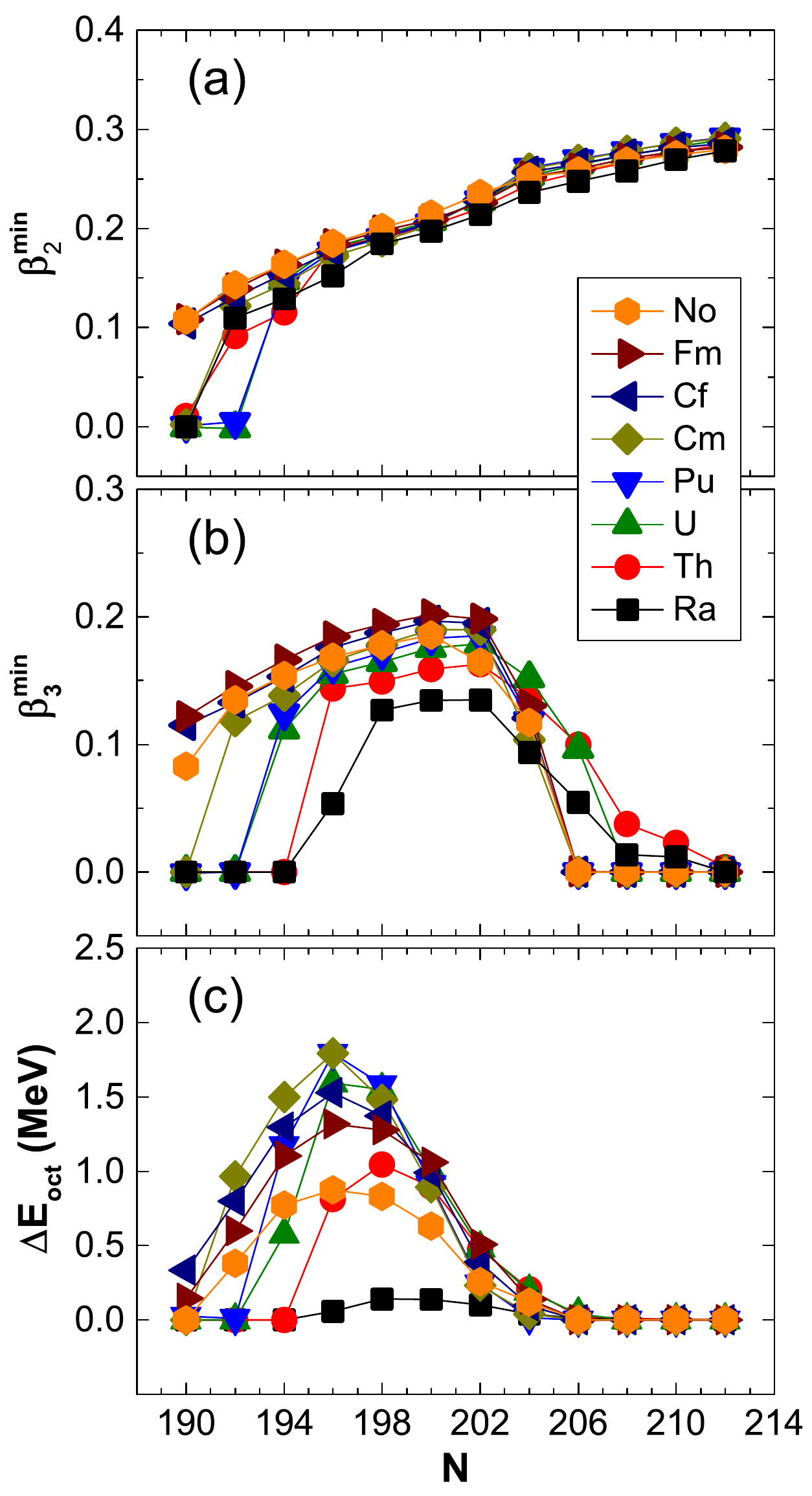}
\figcaption{\label{bmin}  Calculated values of the equilibrium quadrupole $\beta_2$ and octupole $\beta_3$ deformations as well as the octupole deformation energy $\Delta E_\text{oct}$  as functions of the neutron number for the eight isotopic chains analyzed in the present study.}
\end{center}


\begin{center}
\includegraphics[width=7.5cm]{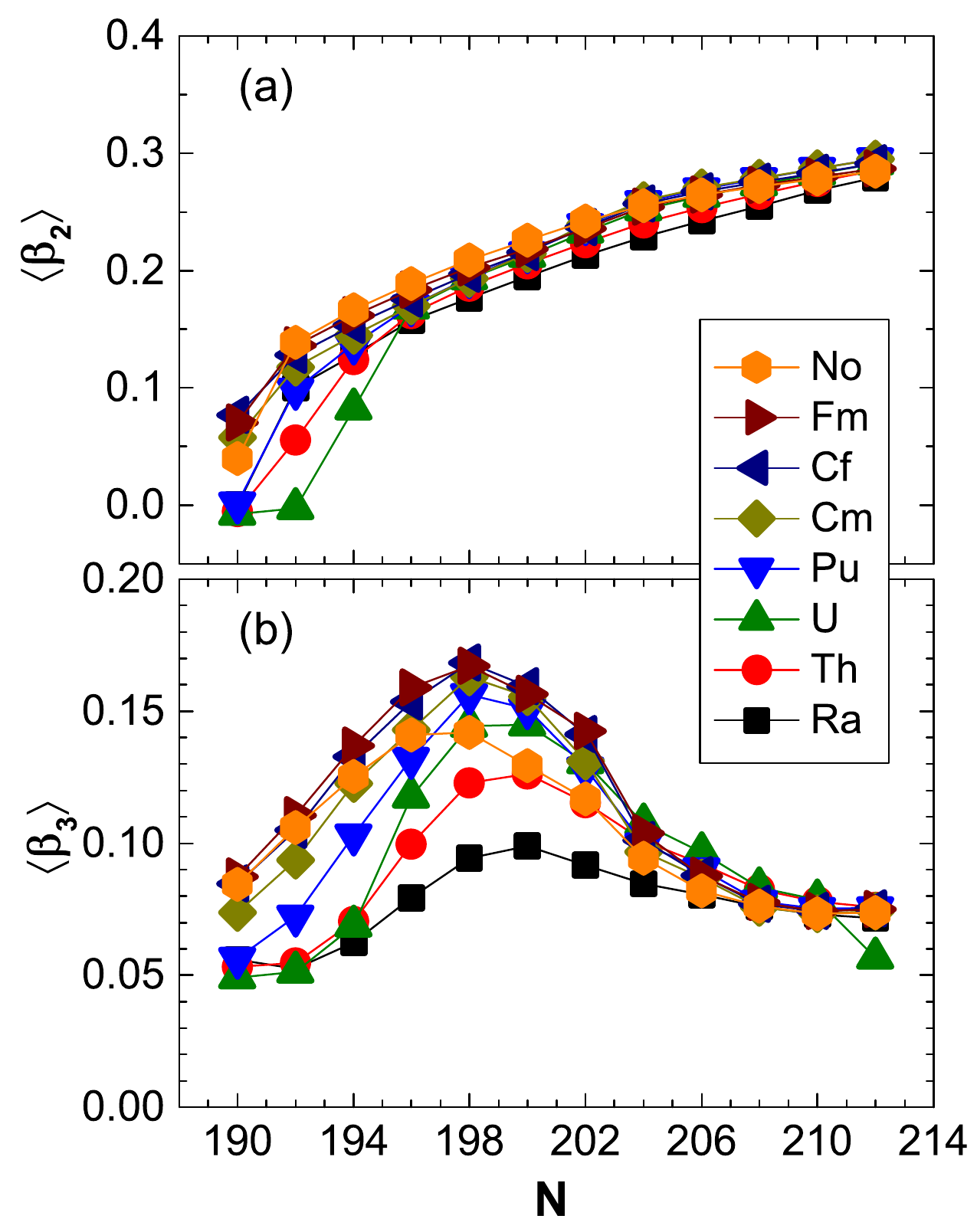}
\figcaption{\label{aveb}  Mean values of the quadrupole $\langle\beta_2\rangle$ and octupole $\langle\beta_3\rangle$ deformations, computed for the QOCH ground states $0^+_1$, as functions of the neutron number.}
\end{center}

Figure \ref{aveb} displays the expectation values of the quadrupole $\langle\beta_2\rangle$ and octupole $\langle\beta_3\rangle$ deformations in the QOCH ground states $0^+_1$ as functions of the neutron number. Initially the ground-state quadrupole deformation $\langle\beta_2\rangle$ increases rapidly up to $N\sim194$ and then more gradually with neutron number.  The corresponding ground-state octupole deformation $\langle\beta_3\rangle$ increases at first, and then decreases, with peaks at $N\sim198$. In our calculation $\langle\beta_3\rangle\gtrsim0.10$ is predicted for octupole-deformed nuclei.

%
\end{multicols}
\ruleup
\begin{center}
\includegraphics[width=17cm]{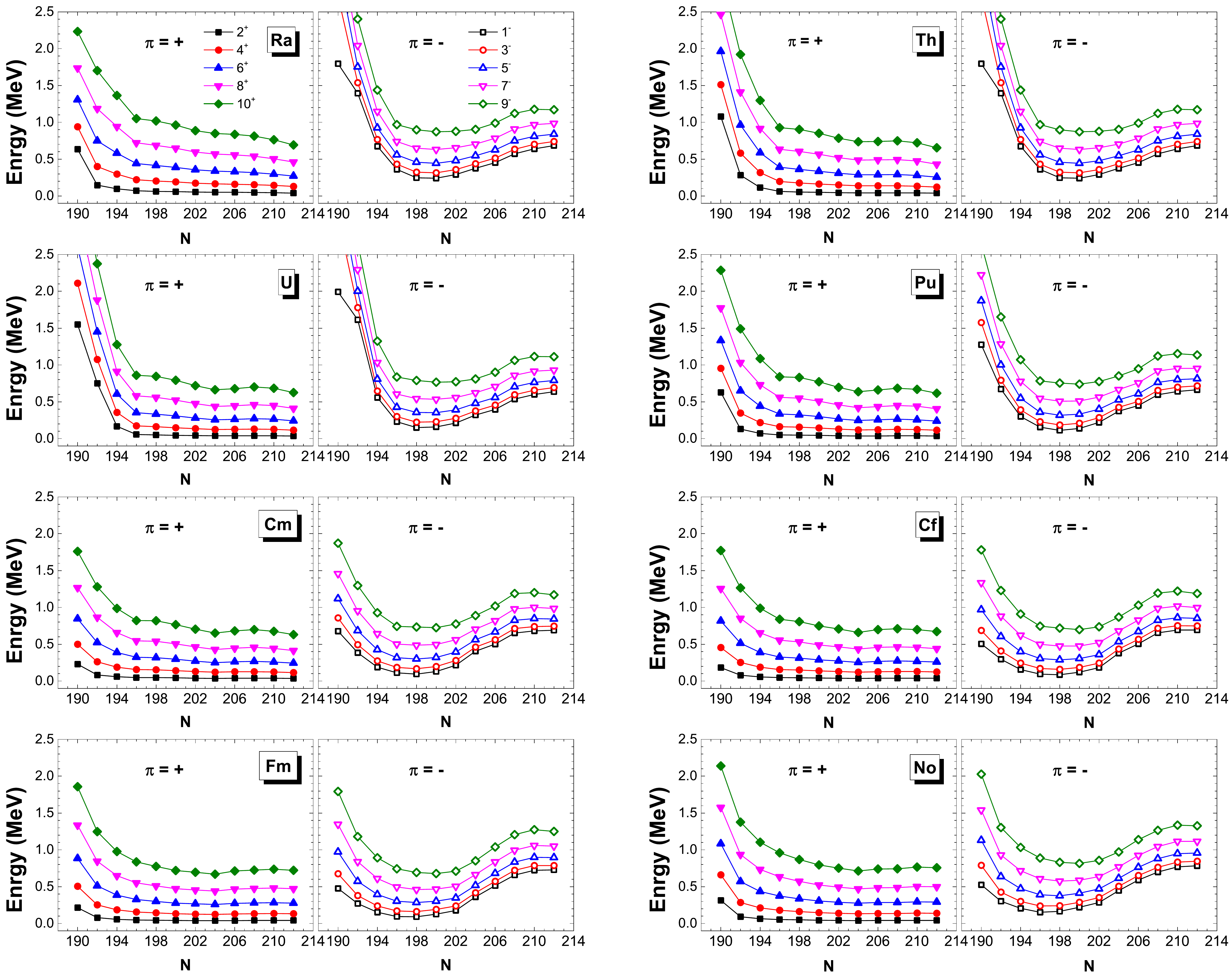}
\figcaption{\label{spec}  The energy spectra of the low-lying even-spin positive-parity states up to $J^\pi=10^+$ and odd-spin negative-parity states up to $J^\pi=9^-$,  as functions of the neutron number, for the eight isotopic chains analyzed in the present study.}
\end{center}
\ruledown

\begin{multicols}{2}

Figure \ref{spec} displays the energy spectra of the lowest-lying even-spin positive-parity states up to $J^{\pi}=10^+$ and odd-spin negative-parity states up to $J^{\pi}=9^-$ for the eight isotopic chains. For the positive-parity bands, the excitation energies drop rapidly until $N\sim196$, and then vary slowly when adding more neutrons.  The excitation energies of the negative-parity bands exhibit a parabolic behavior with minima at $N\sim198$. Specifically, the minima gradually evolve from $N=200$ for Ra and Th isotopes to $N=198$ for U, Pu, Cm, Cf, and Fm isotopes, and finally $N=196$ for No isotopes. The lowest $1^-_1$ state is observed at $^{296}$Cf and the excitation energy is 0.086 MeV. The evolutions of the positive- and negative-parity bands with neutron number are correlated with those of the average quadrupole $\langle\beta_2\rangle$ and octupole $\langle\beta_3\rangle$ deformations in Fig. \ref{aveb}, respectively. The quadrupole deformation reflects the collectivity of a nucleus and, generally larger $\langle\beta_2\rangle$ leads to a more condensed ground state band. On the other hand, the larger octupole deformation corresponds to stronger octupole correlation, which drives the negative-parity band closer to the ground state.

Another indication of the  shape transition, from nearly spherical to octupole-deformed to octupole-soft, is the odd-even staggering shown in Fig. \ref{stagger}. For both positive and negative parity, we plot the calculated ratios $E(J )/E(2^+_1)$ for the yrast states of the eight isotopic chains as functions of the angular momentum $J$.  The ratios of the $N=190$ isotones are almost linear as a function of $J$, indicating that they are nearly spherical nuclei. For the $N\geq202$ isotones,  the odd-even staggering becomes more pronounced, and this means that negative-parity states form a separate rotational-like collective band built on the octupole vibrational state. In between, for the isotopes $^{286-290}$Th, $^{288-292}$U, $^{288-294}$Pu, $^{288-296}$Cm, $^{290-298}$Cf, $^{292-300}$Fm, and $^{294-302}$No, the ratios are parabolic as a function of $J$ and the odd-even staggering is negligible, indicating that positive- and negative-parity states form a single rotational band.  Therefore, these isotopes are stable octupole-deformed. For Ra isotopes, pronounced odd-even staggering is observed in nuclei with $N\geq192$.

\end{multicols}
\ruleup
\begin{center}
\includegraphics[width=15cm]{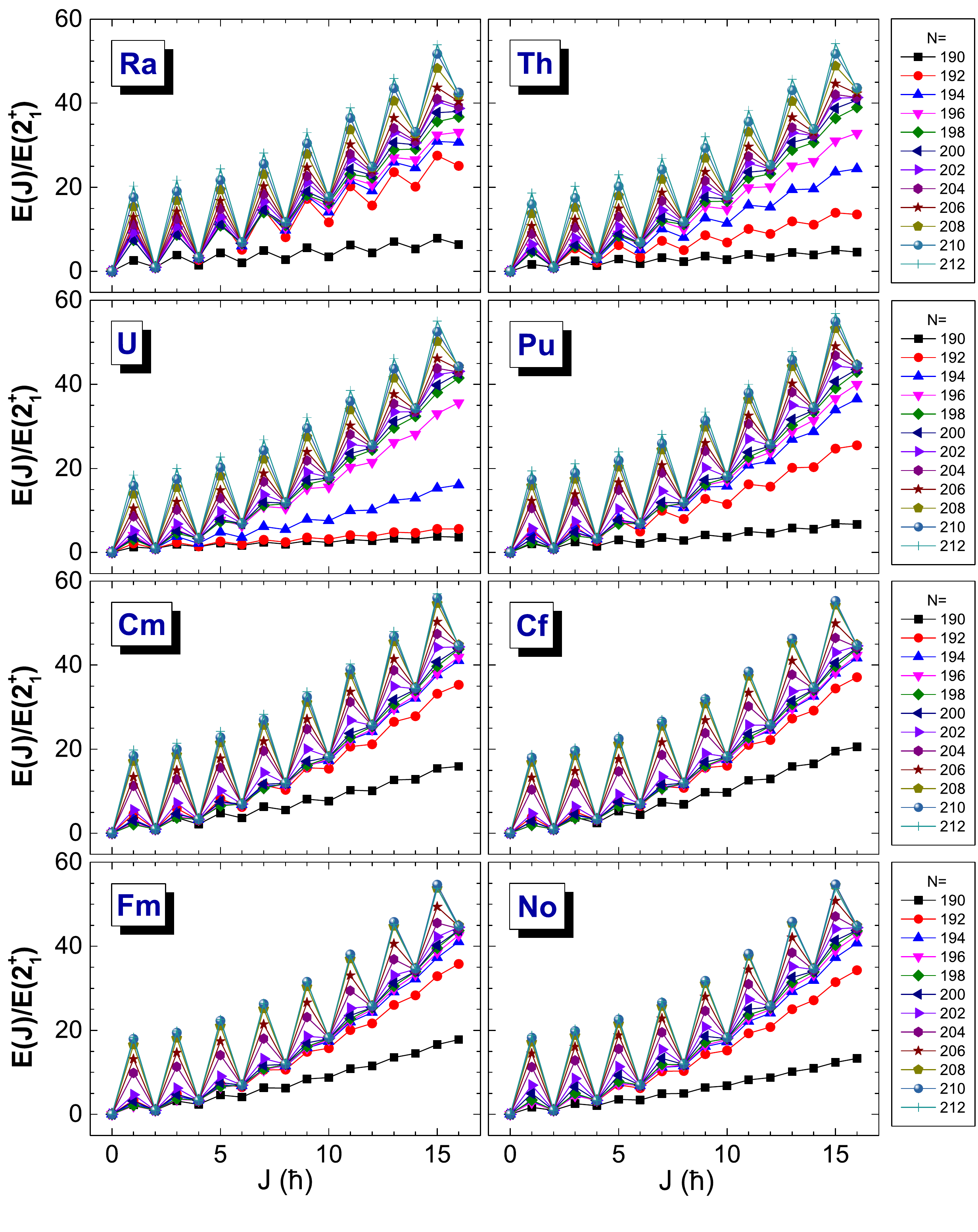}
\figcaption{\label{stagger}  The calculated energy ratios $E(J^\pi)/E(2^+_1)$ for states of the positive-parity ground-state band ($J$ even) and the lowest negative-parity band ($J$ odd) as functions of the angular momentum $J$ for the eight isotopic chains analyzed in the present study.}
\end{center}
\ruledown

\begin{multicols}{2}

\begin{center}
\includegraphics[width=8cm]{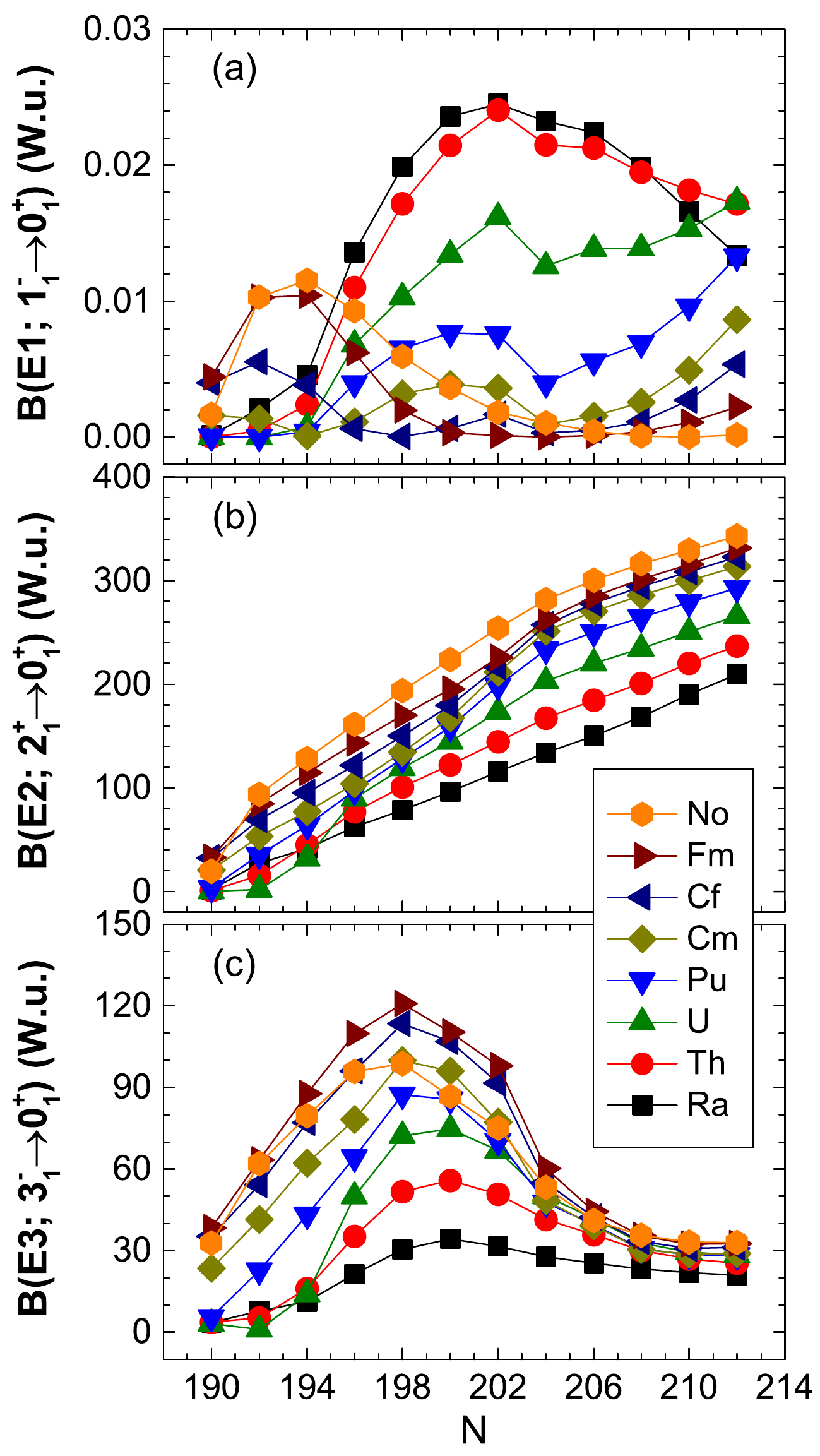}
\figcaption{\label{BE}  The calculated $B(E1; 1^-_1\to 0^+_1)$, $B(E2; 2^+_1\to 0^+_1)$, and $B(E3; 3^-_1\to 0^+_1)$ (in units of W.u.)  as functions of the neutron number for the eight isotopic chains analyzed in the present study.}
\end{center}

Low energy $B(E3)$ values are good measures of octupole collectivity. For low-lying states in nuclei, the ground-state $B(E3)$ transition probabilities present a maximum value in the region of octupole-deformed nuclei \cite{Butler96}. Figure \ref{BE} displays the calculated $B(E1; 1^-_1\to 0^+_1)$, $B(E2; 2^+_1\to 0^+_1)$, and $B(E3; 3^-_1\to 0^+_1)$ (in units of W.u.)  as functions of the neutron number for the eight isotopic chains analyzed in the present study.  The $B(E3; 3^-_1\to 0^+_1)$ increases at first, and then decreases with peaks at $N\sim198$. This is consistent with the evolution of the average octupole deformation $\langle\beta_3\rangle$ (cf. Fig. \ref{aveb}).  In our calculation, $B(E3; 3^-_1\to 0^+_1)>60$ W.u. is predicted for the pronounced octupole-deformed nuclei. Large $B(E1; 1^-_1\to 0^+_1)$ is observed in the heavier Ra, Th, U, and Pu isotopes, and lighter Fm and No isotopes, different from the evolution of $B(E3; 3^-_1\to 0^+_1)$. This may be because the electric dipole moment is not only dependent on the octupole correlation, but also sensitive to the shell effects and occupancy of different orbitals \cite{Butler96,Bucher17}. The $B(E2)$ values increase gradually to more than 200 W.u. with increasing neutron number, indicating a shape transition from nearly spherical to well-deformed shapes for all the isotopic chains.

\begin{center}
\includegraphics[width=8.5cm]{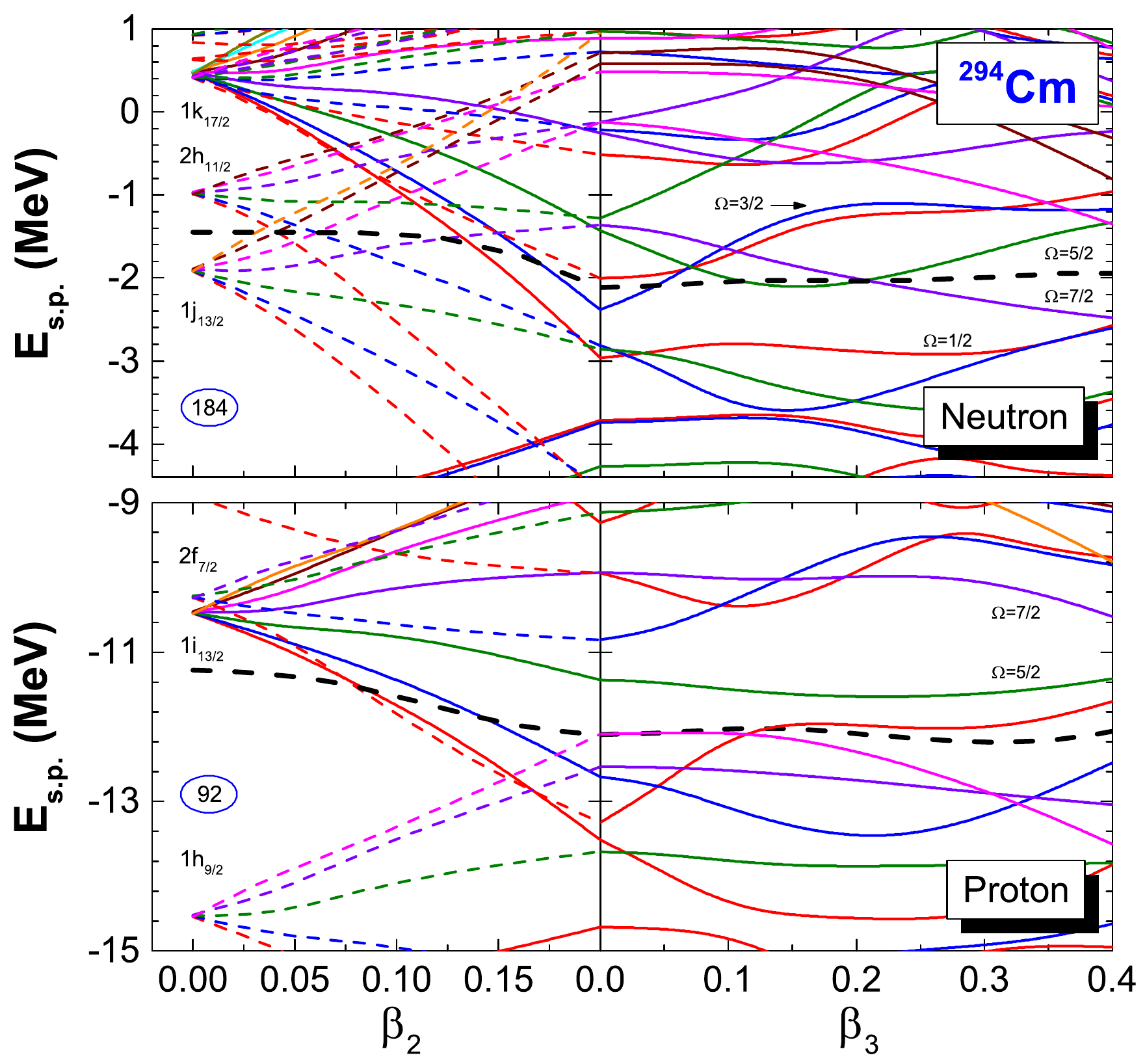}
\figcaption{\label{singlepart}  Single-neutron levels (top panel) and single-proton levels (bottom panel) of $^{294}\text{Cm}$ as functions of the deformation parameters, calculated by the RMF+BCS based on PC-PK1 energy density functional. Each plot follows the quadrupole deformation parameter $\beta_2$ up to the position of the equilibrium minimum $\beta_2=0.20$, with the constant octupole deformation parameter $\beta_3=0$ (left panels). For the constant value $\beta_2=0.20$, the panels on the right display the dependence of the single-nucleon energies on the octupole deformation, from $\beta_3=0$ to $\beta_3=0.4$. The thick dashed (black) curves denote the Fermi levels.}
\end{center}

A microscopic picture of the onset of octupole deformation and octupole softness emerges from the dependence of the single-nucleon levels on the two deformation parameters $\beta_2$ and $\beta_3$. In Fig. \ref{singlepart} we plot the single neutron and proton levels of $^{294}\text{Cm}$ along a path in the $\beta_2$-$\beta_3$ plane, calculated by the RMF+BCS based on PC-PK1 energy density functional.  They are similar to the usual Nilsson orbitals, but we also show their evolution along the octupole direction. In the mean-field approach there is a close relation between the total binding energy and the level density around the Fermi level in the Nilsson diagram of single-particle energies. A lower-than-average density of single-particle levels around the Fermi energy results in extra binding, whereas a larger-than-average value reduces binding. Therefore, the onset of octupole minima around $^{294}\text{Cm}$ (c.f. Figs. \ref{RaTh-pes}-\ref{FmNo-pes}) can be attributed to the low neutron-level density around the Fermi surface at $N\sim198$ and $\beta_3\sim0.20$, induced by the repulsion between the $\Omega=3/2$ pair of levels (blue curves) that originate from the $(h_{11/2}, k_{17/2})$ spherical neutron levels. A low neutron-level density is also predicted at $N\sim204$, which may cause the octupole softness in the isotopes with $N\geq204$. Moreover, it is noted that an octupole-deformed proton shell gap at $Z\sim100$ is obtained, which may enhance the octupole deformations in the heavier isotopic chains (c.f. Figs. \ref{aveb}, \ref{BE}).

%
\section{\label{IV}Summary}

In the present study we have performed a microscopic analysis of octupole shape transitions in eight isotopic chains: Ra, Th, U, Pu, Cm, Cf, Fm, and No with neutron number $190\leq N\leq 212$.  Starting from self-consistent binding energy maps in the $\beta_2$-$\beta_3$ plane, calculated with the RMF+BCS model based on the functional PC-PK1 and $\delta$-force pairing, a recent implementation of the quadrupole-octupole collective Hamiltonian for  vibrations and rotations has been used to calculate the spectroscopy of quadrupole and octupole states of the 96 even-even nuclei. The microscopic deformation energy surfaces exhibit transitions with increasing neutron number: from spherical quadrupole vibrational to stable octupole deformed nuclei, and finally to octupole vibrations characteristic for $\beta_3$-soft potentials in the neutron-rich actinides. The systematics of the energy spectra, odd-even staggering, and transition rates, associated with both positive- and negative-parity yrast states, points to the appearance of prominent octupole correlations around $Z\sim96, N\sim198$, and the corresponding lowering in energy of negative-parity bands with respect to the positive-parity ground-state band.

A microscopic picture of the onset of octupole deformation emerges from the dependence of the single-nucleon levels on the two deformation parameters. The onset of ocutpole minima around $^{294}$Cm can be mainly attributed to the low neutron-level density around the Fermi surface at $N\sim198$ and $\beta_3\sim0.20$, which is induced by the repulsion of the $\Omega=3/2$ level pair originating from the $(h_{11/2}, k_{17/2})$ spherical neutron levels.

\end{multicols}

\clearpage

\end{CJK*}

\vspace{3mm}


\begin{thebibliography}{90}

\vspace{3mm}

%
%


\bibitem{Butler96} P. A. Butler and W. Nazarewicz, Rev. Mod. Phys. {\bf 68}, 349 (1996).

\bibitem{Ahmad93} I. Ahmad and P. A. Butler, Annu. Rev. Nucl. Part. Sci. {\bf 43}, 71 (1993).

\bibitem{BW.15} P. A. Butler and L. Willmann, Nucl. Phys. News {\bf 25}, 12 (2015).

\bibitem{Butler16} P. A. Butler, J. Phys. G {\bf 43}, 073002 (2016).

\bibitem{Gaff13} L. P. Gaffney {\it et al.}, Nature {\bf 497}, 199 (2013).	

\bibitem{Bucher16} B. Bucher {\it et al.}, Phys. Rev. Lett. {\bf 116}, 112503 (2016).

\bibitem{Bucher17} B. Bucher {\it et al.}, Phys. Rev. Lett. {\bf 118}, 152504 (2017).



\bibitem{Bonche86} P. Bonche, P.-H. Heenen, H. Flocard, and D. Vautherin, Phys. Lett. B {\bf 175}, 387 (1986).

\bibitem{Bonche88} P. Bonche, in {\it The Variation of Nuclear Shapes}, edited by J. D. Garrett (World Scientific, Singapore, 1988), p. 302.

\bibitem{Egido91} J. L. Egido and L. M. Robledo, Nucl. Phys. A {\bf 524}, 65 (1991).

\bibitem{Rutz95} K. Rutz, J. A. Maruhn, P. G. Reinhard, and W. Greiner, Nucl. Phys. A 590, 680 (1995).

\bibitem{Geng07} L. S. Geng, J. Meng, and H. Toki, Chin. Phys. Lett. {\bf 24}, 1865 (2007).

\bibitem{Guo10} J.-Y. Guo, P. Jiao, and X.-Z. Fang, Phys. Rev. C {\bf 82}, 047301 (2010).

\bibitem{Robledo10} L. M. Robledo, M. Baldo, P. Schuck, and X. Vi\~{n}as, Phys. Rev. C {\bf 81}, 034315 (2010).

\bibitem{Robledo11} L. M. Robledo and G. F. Bertsch, Phys. Rev. C {\bf 84}, 054302 (2011).

\bibitem{Rodr12} R. Rodr\'{\i}guez-Guzm\'an, L.M. Robledo, and P. Sarriguren, Phys. Rev. C {\bf 86}, 034336 (2012).

\bibitem{Robledo13} L. M. Robledo and P. A. Butler, Phys. Rev. C {\bf 88}, 051302 (2013).

\bibitem{Robledo15} L. M. Robledo, J. Phys. G {\bf 42}, 055109 (2015).

\bibitem{Bernard16} R\'{e}mi N. Bernard, Luis M. Robledo, and Tom\'{a}s R. Rodr\'{\i}guez, Phys. Rev. C {\bf 93}, 061302(R) (2016).


\bibitem{Zhao12} J. Zhao, B.-N. Lu, E.-G. Zhao, and S.-G. Zhou, Phys. Rev. C {\bf 86}, 057304 (2012).

\bibitem{ZhouSG16}  S.-G. Zhou, Phys. Scr. {\bf 91}, 063008 (2016).

\bibitem{Zhao17} J. Zhao, B.-N. Lu, E.-G. Zhao, and S.-G. Zhou, Phys. Rev. C {\bf 95}, 014320 (2017).


\bibitem{Nomura13} K. Nomura, D. Vretenar, and B.-N.Lu, Phys.Rev.C {\bf 88}, 021303 (2013).

\bibitem{Nomura14} K. Nomura, D. Vretenar, T. Nik\v{s}i\'{c}, and B.-N. Lu, Phys. Rev. C {\bf 89}, 024312 (2014).

\bibitem{Nomura15} K. Nomura, R. Rodr\'{\i}guez-Guzm\'an, and L. M. Robledo, Phys. Rev. C {\bf 92}, 014312 (2015).

\bibitem{Agbe16} S. E. Agbemava, A. V. Afanasjev, and P. Ring, Phys. Rev. C {\bf 93}, 044304 (2016).

\bibitem{Agbe17} S. E. Agbemava and A. V. Afanasjev, Phys. Rev. C {\bf 96}, 024301 (2017).

\bibitem{Ebata17} S. Ebata and T. Nakatsukasa, Phys. Scr. {\bf 92}, 064005 (2017).

\bibitem{Zhang10} W. Zhang, Z.-P. Li, and S.-Q.Zhang, Chi. Ph. C {\bf 34}, 1094 (2010).

\bibitem{Zhang10b} W. Zhang, Z. P. Li, S. Q. Zhang, and J.Meng, Phys. Rev. C {\bf 81}, 034302 (2010).

\bibitem{Li13} Z. P. Li, B. Y. Song, J. M. Yao, D. Vretenar, and J. Meng, Phys. Lett. B {\bf 726}, 866 (2013).

\bibitem{Yao15}  J. M. Yao, E. F. Zhou, and Z. P. Li, Phys. Rev. C {\bf 92}, 041304(R) (2015).

\bibitem{Li16} Z. P. Li, T. Nik\v{s}i\'{c}, and D. Vretenar, J. Phys. G {\bf 43}, 024005 (2016).

\bibitem{Zhou16}  E. F. Zhou, J. M. Yao, Z. P. Li, J. Meng, and P. Ring, Phys. Lett. B {\bf 753}, 227 (2016).

\bibitem{Xia17} S. Y. Xia, H. Tao, Y. Lu, Z. P. Li, T. Nik\v{s}i\'{c} and D. Vretenar. Submitted to Phys. Rev. C.



\bibitem{Naza84} W. Nazarewicz, P. Olanders, I. Ragnarsson, J. Dudek, G. A. Leander, P. Moller, and E. Ruchowsa, Nucl. Phys. A {\bf 429}, 269 (1984).

\bibitem{Moller08} P. M\"{o}ller, R. Bengtsson, B. Carlsson, P. Olivius, T. Ichikawa, H. Sagawa, and A. Iwamoto, At. Data Nucl. Data Tables {\bf 94}, 758 (2008).

\bibitem{Wang15} H.-L. Wang, J. Yang, M.-L. Liu, and F.-R. Xu, Phys. Rev. C {\bf 92}, 024303 (2015).



\bibitem{Scho78} O. Scholten, F. Iachello, and A. Arima, Ann. Phys. (NY) {\bf 115}, 325 (1978).

\bibitem{Otsuka88} T. Otsuka and M. Sugita, Phys. Lett. B {\bf 209}, 140 (1988).

\bibitem{Bizzeti04} P. G. Bizzeti and A. M. Bizzeti-Sona, Phys. Rev. C {\bf 70}, 064319 (2004).

\bibitem{Bonatsos05} D. Bonatsos, D. Lenis, N. Minkov, D. Petrellis, and P. Yotov, Phys. Rev. C {\bf 71} (2005) 064309.

\bibitem{Bizzeti13} P. G. Bizzeti and A. M. Bizzeti-Sona, Phys. Rev. C {\bf 88}, 011305(R) (2013).

\bibitem{Minkov12} N. Minkov, S. Drenska, M. Strecker, W. Scheid, and H. Lenske, Phys. Rev. C {\bf 85}, 034306 (2012).

\bibitem{Jolos12} R. V. Jolos, P. von Brentano, and J. Jolie, Phys. Rev. C {\bf 86}, 024319 (2012).


\bibitem{Chen15} Y.-J. Chen, Z.-C. Gao, Y.-S. Chen, and Y. Tu, Phys. Rev. C {\bf 91}, 014317 (2015).




\bibitem{Bender03} M. Bender, P.-H. Heenen, and P.-G. Reinhard, Rev. Mod. Phys. {\bf 75}, 121 (2003).

\bibitem {VALR05}D. Vretenar, A. V. Afanasjev, G. A. Lalazissis, and P. Ring,
    Phys. Rep. {\bf 409}, 101 (2005).

\bibitem{Meng06} J. Meng, H. Toki, S. G. Zhou, S. Q. Zhang, W. H. Long, and L. S. Geng,
    Prog. Part. Nucl. Phys. {\bf 57}, 470 (2006).

\bibitem{Stone07} J. Stone and P.-G. Reinhard, Prog. Part. Nucl. Phys. {\bf 58}, 587 (2007).

\bibitem{Nik11} T. Nik\v{s}i\'{c}, D. Vretenar, and P. Ring, Prog. Part. Nucl. Phys. {\bf 66}, 519 (2011).

\bibitem{Meng16} {\it Relativistic Density Functional for Nuclear Structure}, edited by J. Meng (World Scientic, Singapore, 2016).



\bibitem{Ring80} P. Ring and P. Schuck, {\it The Nuclear Many-Body Problem} (Springer-Verlag, Heidelberg, 1980).

\bibitem{Yao14} J. M. Yao, K. Hagino, Z. P. Li, J. Meng, and P. Ring, Phys. Rev. C  {\bf 89}, 054306 (2014).

\bibitem{Zhao10} P. W. Zhao, Z. P. Li, J. M. Yao, and J. Meng, Phys. Rev. C {\bf 82}, 054319 (2010).






\bibitem{Zhang14} Q. S. Zhang, Z. M. Niu, Z. P. Li, J. M. Yao,  and J. Meng, Front. Phys. {\bf 9}, 529 (2014).

\bibitem{Lu15} K. Q. Lu, Z. X. Li, Z. P. Li, J. M. Yao,  and J. Meng, Phys. Rev. C {\bf 91}, 027304 (2015).

\bibitem{Quan17} S. Quan, Q. Chen, Z. P. Li, T. Nik\v{s}i\'{c}, and D. Vretenar, Phys. Rev. C {\bf 95}, 054321 (2017).

\bibitem{Inglis56} D. R. Inglis, Phys.Rev. {\bf 103}, 1786 (1956).

\bibitem{Belyaev61} S. T. Belyaev, Nucl. Phys. {\bf 24}, 322 (1961).

\bibitem{Girod79} M. Girod and B. Grammaticos, Nucl. Phys. A {\bf 330}, 40 (1979).

\bibitem{Bender00} M. Bender, K. Rutz, P.-G. Reinhard, and J. A. Maruhn, Eur. Phys. J. A 8, 59 (2000).

\bibitem{Xiang13} J. Xiang, Z. P. Li, J. M. Yao, W. H. Long, P. Ring, and J. Meng, Phys. Rev. C {\bf 88}, 057301 (2013).

\bibitem{Nik14} T. Nik\v{s}i\'{c}, D. Vretenar, and P. Ring,  Comp. Phys. Comm. {\bf 185},  1808 (2014).




%
%
%


\end{thebibliography}
\end{document}